\documentclass[11pt,twoside]{article}

\usepackage{asp2006}
\usepackage{epsf}
\usepackage{psfig}
\usepackage{lscape}

\begin{document}
\title{Galactic Planetary Nebulae in the AKARI Far-Infrared Surveyor Bright Source Catalog}   %%% Fill in title

\author{Nick Cox,$^1$ Arturo Manchado,$^{2,3}$ Pedro Garc\'ia-Lario,$^1$ Ryszard Szczerba$^4$}   %%% Fill in author names
\affil{
$^1$ Herschel Science Centre, European Space Astronomy Centre, European Space Agency, Spain\\
$^2$ Instituto Astrof\'isica de Canarias, Spain\\
$^3$ CSIC, Spain\\
$^4$ N. Copernicus Astronomical Center, Poland
}    %%% Fill in author affiliations

\begin{abstract} 
We present the results of our preliminary study of all known Galactic PNe
(included in the Kerber 2003 catalog) which are detected by the AKARI/FIS All-Sky
Survey as identified in the AKARI/FIS Bright Source \mbox{Catalog} (BSC) Version $\beta$-1.
\end{abstract}

\section{Introduction}
Planetary Nebulae (PNe) are low-mass stars that have evolved along the evolutionary track from AGB, post-AGB,
proto-PN to PN. One important goal of far-infrared (FIR) studies of PNe is to identify the FIR excess which provides
important information on the mass loss history experienced by these stars in the recent past.
We have performed a preliminary analysis of known Galactic PNe (taken from the \citet{Kerber2003} catalog) and their
presence and properties in the $\beta$-1 release of the AKARI/FIS bright source \mbox{catalog} (BSC). 
We discuss potential technical issues related to the catalog and the effect on the study of PNe. Independent AKARI and IRAS spectral energy distributions
(SEDs) are fitted with simple dust models to understand detection limits, calibration limitations and biases affecting
the PN sample.

\section{Galactic PNe in the BSC}
About 40\% of the known PNe (\citealt{Kerber2003}) are detected by AKARI/FIS while 63\% can be identified in the IRAS point
source catalog (PSC). This discrepancy is partly explained by gaps in sky-coverage of the AKARI/FIS BSC but also by
differences in the point source flux sensitivity between these two surveys.
Currently, there is not yet information available on the sky coverage (e.g.~number of scans across a certain region of
the sky). Noteworthy is the fact that AKARI/FIS detected a number of PNe that were not detected by IRAS.

\begin{figure}[ht!]
\begin{minipage}[b]{0.45\linewidth}
   \centering
   \resizebox{1\hsize}{!}{
      \includegraphics{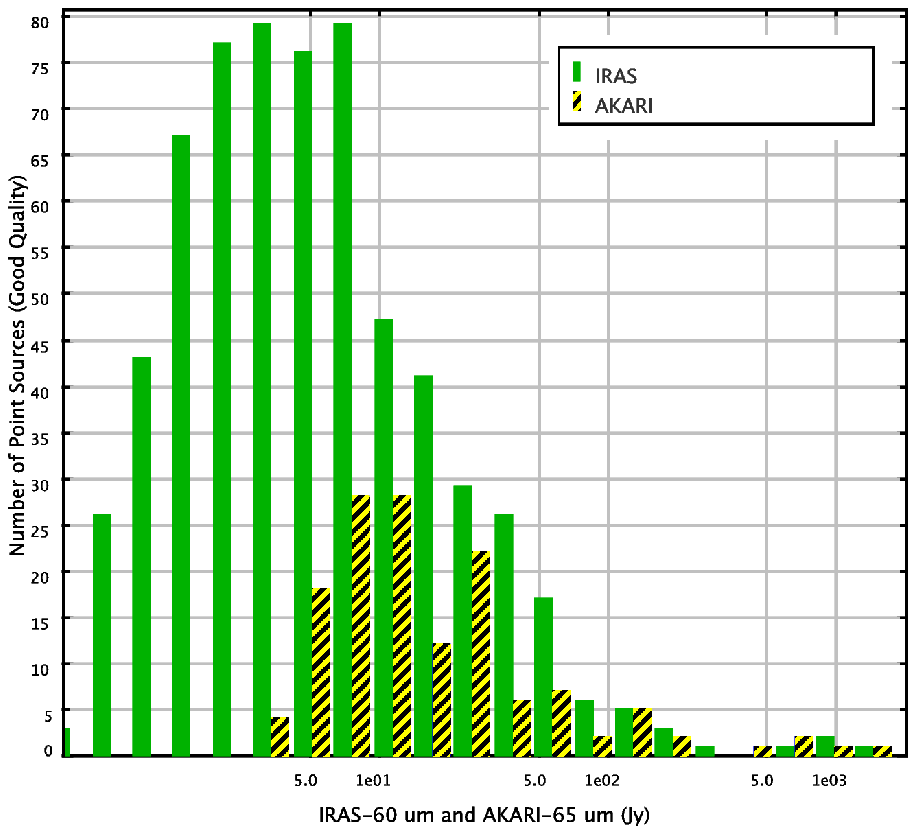}
   }
   \caption{PNe detected by IRAS-60~$\mu$m and AKARI-65~$\mu$m bands.}\label{fig:fp}
\end{minipage}
\hspace{0.3cm}
\begin{minipage}[b]{0.45\linewidth}
   \centering
   \resizebox{1\hsize}{!}{
      \includegraphics{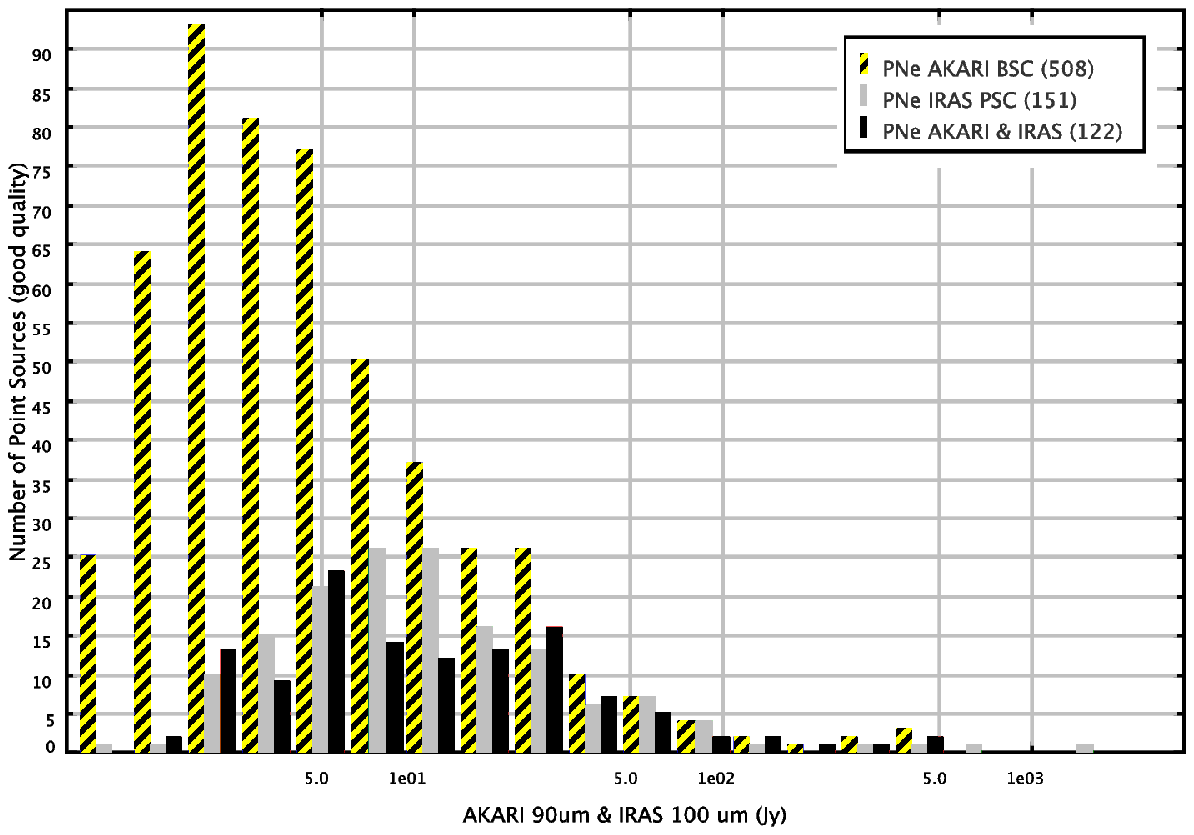}
   }
   \caption{PNe detected by AKARI-90~$\mu$m, IRAS-100~$\mu$m bands, and both.}\label{fig:ldl}
\end{minipage}

\end{figure}

\subsection{Flux Sensitivity and Completeness}
From the initial set of PNe found in both IRAS and AKARI catalogs we selected only those with good quality flux values
at 60 and 65~$\mu$m, respectively. We selected this band since the IRAS-100 fluxes are generally of worse quality
though the AKARI-90 fluxes are better than the AKARI-65 fluxes (only 139 have good quality 65 flux; and 629 good
IRAS-60).
For the AKARI-65 band there are (almost) no reliable point sources with flux densities less than 4 Jy, whereas IRAS
detected many PNe below the 1~Jy limit.
This can be clearly seen from Figure 1 which shows the number of sources (of good quality) versus the IRAS and
AKARI fluxes.
The picture for the AKARI-90 and IRAS-100 fluxes is, however, the opposite.
For this wavelength IRAS is severely limited (by presence of cirrus emission) whereas AKARI gives many good
detections (also below 4~Jy). See Figure~2.

\subsection{Absolute Flux Calibration}
The IRAS-60 and AKARI-65 fluxes are, as expected for PNe, roughly linearly correlated. 
The exact flux-ratio between these two bands depends on the colour temperature at
which these sources peak (ie. the slope between 60 and 65~$\mu$m changes with dust temperature).
In Figure 3 we plot the IRAS-60/AKARI-65 ratio versus AKARI-65 flux (linear y-scale and log
x-scale). All PNe (found in both AKARI \& IRAS) are plotted in grey, and those with \mbox{quality-3} 
data in both bands in black. The correlation between IRAS and AKARI fluxes is much better for 
the latter ``quality'' (green) selected set.
Sources with black-body temperature 80-100~K peak at wavelengths less than 60~$\mu$m, and therefore, 
as expected, the IRAS-60 fluxes are somewhat higher than the AKARI-65 fluxes. Though based on 
the SEDs the difference is larger than expected (see below).
For the AKARI-90 and IRAS-100~$\mu$m the result is similar. 
In this case we have created a subset from all cross-matches (grey) to include only those targets with quality fluxes
and with cirrus emission less than 100 MJy/sr (black). This greatly improves the correlation. See Figure~4.

\begin{figure}[th!]
\begin{minipage}[t]{0.45\linewidth}
   \centering
   \resizebox{!}{5.5cm}{\includegraphics{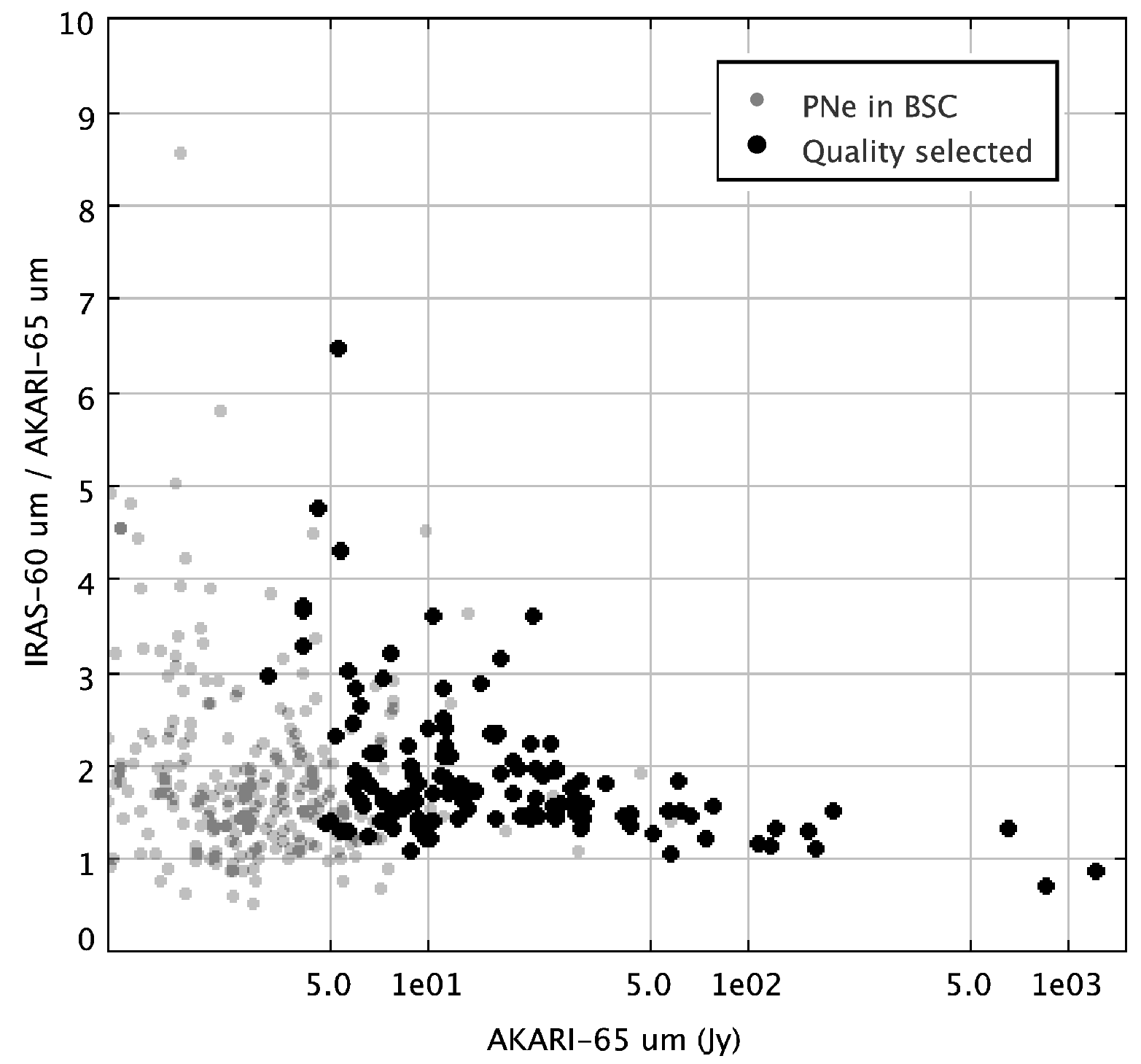}}
   \caption{IRAS-60/AKARI-65 flux ratio versus AKARI-65 flux (\mbox{linear} y-scale and log x-scale).
Good quality data (quality-3) are those in black.}\label{fig:3}
\end{minipage}
\hspace{0.2cm}
\begin{minipage}[t]{0.45\linewidth}
   \centering
   \resizebox{!}{5.5cm}{ \includegraphics{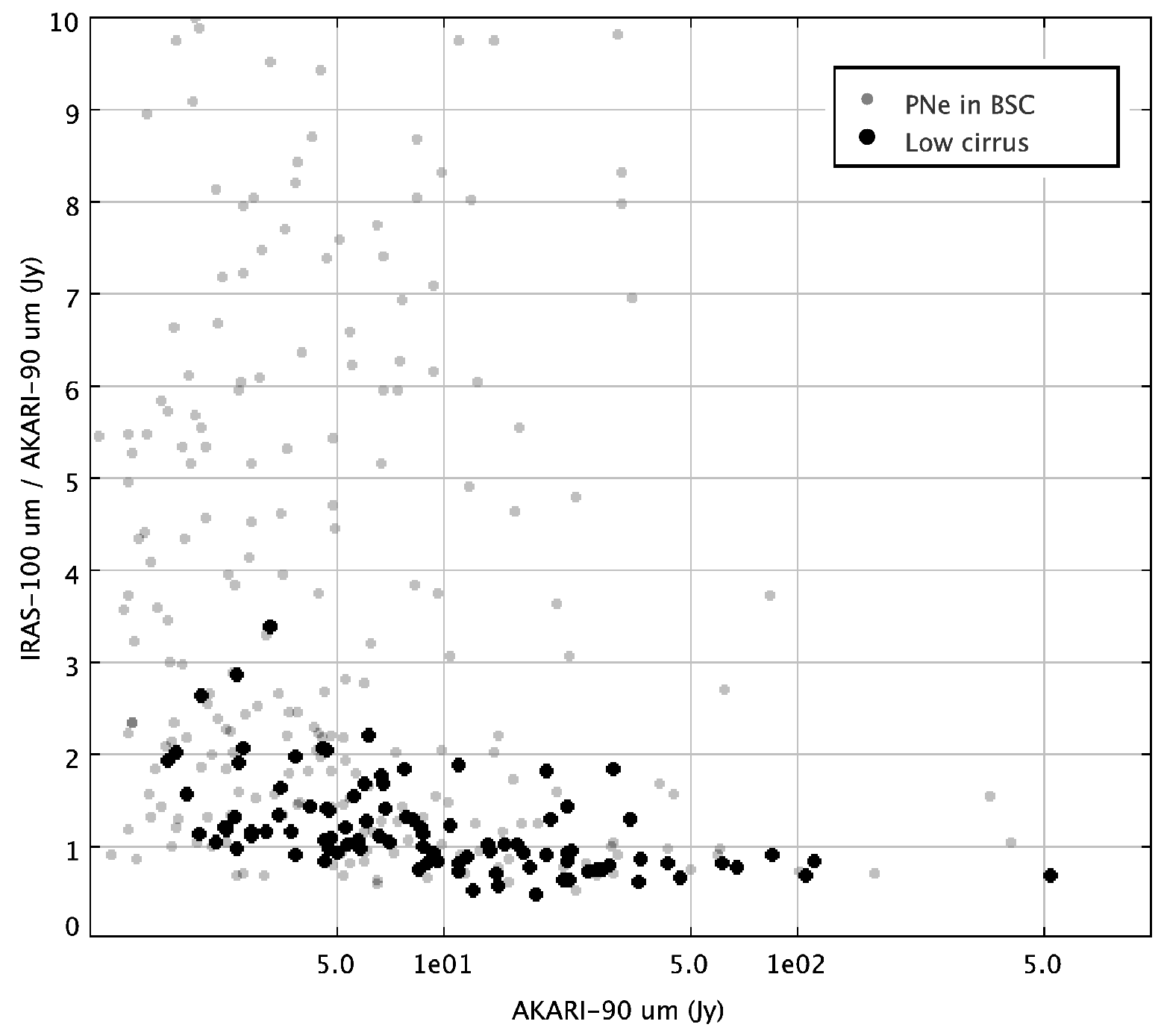}
   }
   \caption{IRAS-100/AKARI-90 flux ratio vs. AKARI-90 flux (linear y-scale and log x-scale). Good quality data
and cirrus emission less than 100 MJy/sr are those in black.}\label{fig:4}
\end{minipage}

\end{figure}

\begin{table}[!hb]
\begin{center}
\caption{Dust temperatures for PNe detected in all IRAS and AKARI bands.}\label{tb:dusttemp}
\smallskip
{\small
\begin{tabular}{lll}\hline
PN	      & T$_{\rm AKARI}$ & T$_{\rm 12-140}$ \\ \hline
A30	      & 46.8	      &      50.0      \\
CRL618	      & 98.0	      &      78.0      \\
PN G009.3+05.7& 54.5	      &      61.8      \\
He2-113	      & 60.7	      &      89.0      \\
HE3-1333      & 52.8	      &      77.4      \\
IC4406	      & 27.0	      &      49.2      \\
M1-78	      & 53.8	      &      66.0      \\
M2-9	      & 42.7	      &      70.9      \\
NGC 3132      & 30.3	      &      49.0      \\
NGC 6302      & 45.4	      &      64.0      \\
NGC6369	      & 54.1	      &      70.1      \\
NGC 6572      & 58.7	      &      97.6      \\
NGC 6720      & 31.3	      &      53.8      \\
Vo3	      & 27.2	      &      31.8      \\
\hline
\end{tabular}
}
\end{center}
\end{table}

\section{SEDs and Dust Temperatures of the Selected AKARI PNe}
There are (only) 14 PNe that are detected in all AKARI/FIS and IRAS bands. 
We use only these PNe to make a detailed study of the SED.
From fitting the AKARI flux data with one-component black-body we find dust temperatures 
that are systematically lower ($\sim$25\%)
than those derived for the same targets from IRAS fluxes alone.
Table~\ref{tb:dusttemp} list the dust temperatures fitting a single black-body using AKARI data (column 2), 
and IRAS 12, 25 and 60~$\mu$m and AKARI 90 and 140~$\mu$m fluxes (column 3).
For all 14 PNe the IRAS$+$AKARI SED is shown in Figure~5. Also indicated are the cirrus emission values from IRAS.
We note that in particular the 65~$\mu$m flux seems to be often under-estimated for low-medium ($<$100 Jy) flux densities
and may point to systematic errors in the current $\beta$ release of the BSC.
The difference in the derived T$_{\rm dust}$ for both cases arises from taking into account the far-IR 
140 and 160~$\mu$m fluxes in addition to the more accurate 90~$\mu$m flux.

\begin{figure}[ht!]
\begin{center}
   \resizebox{0.8\hsize}{!}{\includegraphics{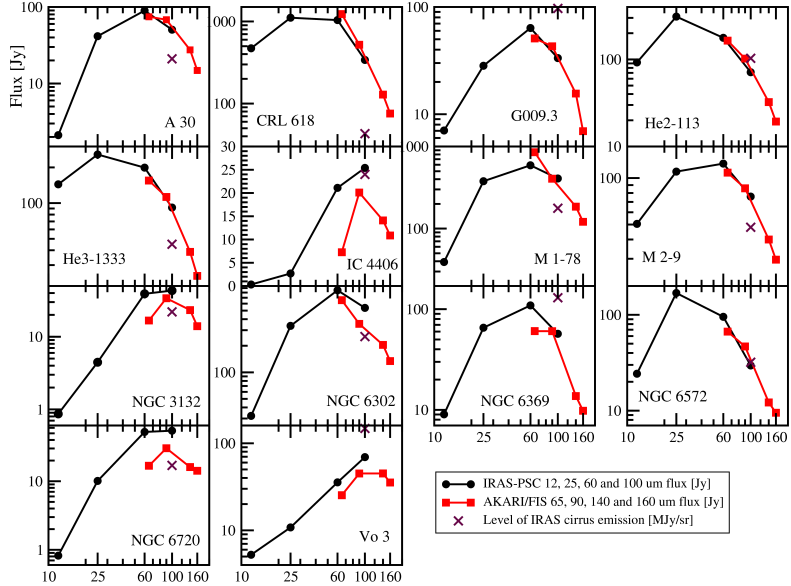}}
\end{center}
\caption{IRAS+AKARI spectral energy distribution of the 14 PNe detected in all bands.
The cirrus emission values from IRAS (at 100~$\mu$m in MJy/sr) are also indicates (crosses).}
\label{fig:}
\end{figure}

\section{Summary and Future Work}
AKARI performs significantly better than IRAS in the 90/100~$\mu$m band.
This, in combinations with its further far-infrared coverage, makes it very 
suitable to detect colder objects (in temperature range from 20 to 70~K).
Future work:
Several black bodies could be fitted to search for different dust components. AKARI plus IRAS
colours could be used to search for new PNe and proto-PNe.

\acknowledgements
We thank the organizers for their hospitality and excellent meeting.
P.G-L and R.Sz. acknowledge support from the Science Faculty of the European Space
Astronomy Center (ESAC).
This research is based on observations with AKARI, a JAXA project with the participation of ESA.

%%% THE BIBLIOGRAPHY
%%%
%%% CONSULT SECTION 3 OF "INSTRUCTIONS FOR AUTHORS" FOR HOW TO USE NATBIB.
%%% AUTHORS ARE ENCOURAGED TO USE EITHER THE "THEBIBLIOGRAPY" ENVIRONMENT
%%% BY UNCOMMENTING (DELETING THE "%" SYMBOL) THE COMMANDS BELOW, OR BY
%%% USING THE BIBTEX ENVIRONMENT. TO FIND OUT WHICH IS APPLICABLE TO YOUR
%%% CONTRIBUTION, CONSULT THE VOLUME EDITORS FOR YOUR PROCEEDINGS.
%%%

\end{document}